\documentclass[12pt]{article}
\usepackage{amsfonts}
\usepackage{amsmath}
\usepackage{epsfig}

\begin{document}

\hfill{revised version 11/2009}\\[-0.9cm]
\vspace{40pt}
\begin{center}
{\Large Cosmological Implications of the Tetron Model 
of Elementary Particles}\\
\vspace{60pt}
{\bf  Bodo Lampe} \\
\vspace{20pt}
\vspace{60pt}

{\bf Abstract}\\
\end{center}
Based on a possible solution to the tetron spin problem, 
a modification of the standard Big Bang scenario is suggested, 
where the advent of a spacetime manifold is connected to 
the appearance of tetronic bound states. The metric tensor 
is constructed from tetron constituents and the reason for 
cosmic inflation is elucidated.  
Furthermore, there are natural dark matter candidates 
in the tetron model. The ratio of ordinary to dark matter in the 
universe is calculated to be 1:5.

\newpage

\section{Introduction} 

Particle physics phenomena can be described, for 
example, by the left-right symmetric Standard Model with gauge group 
$U(1)_{B-L}\times SU(3)_c \times SU(2)_L \times SU(2)_R$ \cite{su2su2} 
and 24 left-handed and 24 right-handed fermion fields 
which including antiparticles amounts to 96 degrees of freedom, 
i.e. this model has right handed neutrinos as well as 
righthanded weak interactions. 

In recent papers \cite{lampe1,lampe2,lampe3} a 
new ordering scheme for the observed 
spectrum of quarks and leptons was presented, which relies on the 
structure of the group of permutations $S_4$ of four objects 
called tetrons, and a mechanism was proposed, how 'germs' of 
the Standard Model interactions might be buried in the  
representations $A_1$, $A_2$, $E$, $T_1$ and $T_2$ 
of this group. Furthermore, it was shown how to 
construct the Standard Model gauge fields with the help of tetrons. 

In the present paper I will argue that this model is not just 
a strange observation in the realm of particle physics, but 
has a more fundamental meaning, so that also gravitational and 
astrophysical effects can also be understood on the tetron basis. 

In modern cosmology there are 3 outstanding phenomena 
not completely understood: the underlying reason for inflation, 
the ratio of dark to ordinary matter and the appearance of 
dark energy: 
\begin{itemize}
\item[i) ] Cosmic inflation \cite{guth} is the widely 
accepted hypothesis that the nascent universe 
passed through a phase of exponential expansion that was driven 
by a vacuum energy density of negative pressure. 
It resolves several problems in the Big Bang cosmology 
that were pointed out in the 1970s, like the horizon 
problem, the flatness problem and the magnetic monopole 
problem. 
\item[ii) ] Dark matter is defined to interact with ordinary matter 
essentially only via gravity.  
Gravitational effects in the rotation of galaxies 
as well as other observations (see e.g. \cite{freeman}) 
suggest the existence of dark matter 
with an amount 4 or 5 times larger than ordinary 
matter which appears in stars, dust and gas. 
\item[iii) ] The present universe is appearantly undergoing 
a phase of accelerated expansion (see e.g. \cite{nesseris}). 
This can be explained either by a modification of the 
Einstein Lagrangian, the so called F(R) gravities, see \cite{cline} 
and references therein, or by the presence of dark energy, 
see e.g. \cite{peebles}, either in the form of a positive cosmological 
constant or of a scalar field, sometimes called 'quintessence' 
\cite{wetterich}, that drives the accelaration 
and acts not unlike the 'inflaton' which is often 
introduced to drive inflation. 
\end{itemize}

In the present paper I want to analyze these phenomena 
in the light of the tetron model. Tetron interactions 
will be assumed to describe the deepest level of matter, 
just above the Planck scale. 
I will show how 
\begin{itemize}
\item[i) ] the tetron model may affect the inflationary 
scenario via the enormous energies set free when after the era 
of a tetron plasma tetron bound states are formed. 
\item[ii) ] some tetron bound states naturally contribute to the 
dark matter of the universe. 
\item[iii) ] tetron interactions may be related to the formation 
of spacetime and the appearance of gravitational forces and 
of dark energy (in the form of a quintessence field).
\end{itemize}

The outline of the article is as follows: in section 2
the main ingredients of the tetron model are reviewed. 
Sections 3, 4 and 6 contain improved arguments as to how 
the spin-$\frac{1}{2}$ properties of quarks and leptons 
can be obtained in this model. In section 5 
the dark matter candidates of the tetron model are 
discussed. In section 7 a view on gravitational interactions 
and dark energy is taken from the standpoint of the tetron model. 
In section 8 I will discuss how shortly after the big bang a 
tetron plasma appears from which in a process of supercooling 
the ordinary quarks, leptons and gauge bosons arise. 
Finally, in the appendix I present an alternative description 
of the tetron idea by introducing both an inner symmetry lattice 
and a spatial lattice. This possibility is related 
to the fact that the permutation group $S_4$ is 
isomorphic to the symmetry group of a tetrahedral lattice. 
Although phenomenologically this approach leads to the same results 
as before, the microscopic interpretation is different 
because tetron bound states are now interpreted 
as lattice excitations of a yet unknown dynamics. 
Sections 5, 7 and 8 and the appendix contain completely 
new material which have not appeared elsewhere. 


\section{Short Review of the Tetron Idea} 

The starting point of refs. \cite{lampe1,lampe2,lampe3} was 
the observation that 
there is a natural one-to-one correspondence between the 
quarks and leptons and the elements of the permutation group 
$S_4$, as made explicit in table 1 and natural in 
the sense that the color, isospin and 
family structure correspond to the K, $Z_2$ 
and $Z_3$ subgroups of $S_4$, where 
$Z_n$ is the cyclic group of n elements and K is the 
so-called Kleinsche Vierergruppe which 
consists of the 3 even permutations $\overline{2143}$, 
$\overline{3412}$, $\overline{4321}$, where 2 pairs of 
numbers are interchanged, plus the 
identity. Note that permutations $\sigma \in S_4$ will be denoted 
$\overline{abcd}, a,b,c,d \in \{1,2,3,4\}$. 

$S_4$ is a semi-direct product 
$S_4= K \diamond Z_3 \diamond Z_2$ where the 
$Z_3$ factor is the family symmetry and $Z_2$ and K can 
be considered to be the 'germs' of weak isospin and color 
symmetry (cf. \cite{lampe2}). 
At low energies this product cannot be distinguished 
from the direct product $K \times Z_3 \times Z_2$ 
but has the advantage of being a simple group and having a 
rich geometric and group theoretical interpretation as 
the rotational symmetry group of a regular tetrahedron and, 
up to a parity factor, the symmetry group of a 
3-dimensional cubic lattice. 
Furthermore it does not only describe quarks and leptons (table 1) 
but also leads to a new ordering scheme for the Standard Model 
(plus some GUT-like) vector bosons, cf. table 2 and ref. \cite{lampe1}. 
In fact, 12 GUT-like heavy vector bosons can be constructed in 
the tetron model, which behave similar, though not identical, 
as the ones appearing in the standard SU(5) model. 

Actually, the assignments in table 1 are only heuristic. 
Instead one has to take linear combinations of symmetry 
adapted wave functions, dictated 
by the 5 representations $A_1$, $A_2$, $E$, $T_1$ and $T_2$ 
of $S_4$ \cite{lampe2}. The content of table 1 may 
then be interpreted as the sum of representations 
$A_1 + A_2 +2E+3T_1+3T_2$.

\begin{table}
\label{tab2}
\begin{center}
\begin{tabular}{|l|c|c|c|}
\hline
&...1234...&...1423...&...1243... \\
& family 1 & family 2 & family 3 \\
\hline
& $\tau$, $b_{1,2,3}$ & $\mu$, $s_{1,2,3}$ & $e$, $d_{1,2,3}$ \\
\hline
$\nu$ & $\overline{1234} (id)$ & $\overline{2314}$ & $\overline{3124}$ \\
$u_1$ & $\overline{2143} (k_1)$ & $\overline{3241}$ & $\overline{1342}$ \\
$u_2$ & $\overline{3412} (k_2)$ & $\overline{1423}$ & $\overline{2431}$ \\
$u_3$ & $\overline{4321} (k_3)$ & $\overline{4132}$ & $\overline{4213}$ \\
\hline
& $\nu_\tau$, $t_{1,2,3}$ & $\nu_{\mu}$, $c_{1,2,3}$  & $\nu_e$, $u_{1,2,3}$  \\
\hline
$l$ & $\overline{3214} (1\leftrightarrow 3)$ & $\overline{1324} (2\leftrightarrow 3)$ & $\overline{2134} (1\leftrightarrow 2)$ \\
$d_1$ & $\overline{2341}$ & $\overline{3142}$ & $\overline{1243} (3\leftrightarrow 4)$ \\
$d_2$ & $\overline{1432} (2\leftrightarrow 4)$ & $\overline{2413}$ & $\overline{3421}$ \\
$d_3$ & $\overline{4123} $ & $\overline{4231} (1\leftrightarrow 4)$ & $\overline{4312}$ \\
\hline
\end{tabular}
\bigskip
\caption{List of elements of $S_4$ ordered in 3 fermion families. 
$k_i$ denote the elements of K and $(a\leftrightarrow b)$ a 
simple permutation where a and b are interchanged. 
Permutations with a 4 at the last position form 
a $S_3$ subgroup of $S_4$ and may be thought of giving the set 
of lepton states. 
It should be noted that this is only a heuristic 
assignment. Actually one has to consider linear combinations 
of permutation states as discussed in section 2.}
\end{center}
\end{table}

\begin{table}
\label{tab7}
\begin{center}
\begin{tabular}{|l|l|l|}
\hline
 $B_{\mu}=\overline{1234} (id)$   & $G_{3\mu}=\overline{2314}$ & $G_{8\mu}=\overline{3124}$ \\
 $W_{3\mu}=\overline{2143} (k_1)$ & $G_{1\mu}=\overline{3241}$ & $G_{2\mu}=\overline{1342}$ \\
 $W_{1\mu}=\overline{3412} (k_2)$ & $G_{4\mu}=\overline{1423}$ & $G_{5\mu}=\overline{2431}$ \\
 $W_{2\mu}=\overline{4321} (k_3)$ & $G_{6\mu}=\overline{4132}$ & $G_{7\mu}=\overline{4213}$ \\
\hline
 $X_{1\mu}=\overline{3214} (1\leftrightarrow 3)$ & $X_{4\mu}=\overline{1324} (2\leftrightarrow 3)$ & $X_{5\mu}=\overline{2134} (1\leftrightarrow 2)$ \\
 $Y_{1\mu}=\overline{2341}$ & $Y_{2\mu}=\overline{3142}$ & $Y_{3\mu}=\overline{1243} (3\leftrightarrow 4)$ \\
 $X_{2\mu}=\overline{1432} (2\leftrightarrow 4)$ & $Y_{4\mu}=\overline{2413}$ & $Y_{5\mu}=\overline{3421}$ \\
 $X_{3\mu}=\overline{4123} $ & $X_{6\mu}=\overline{4231} (1\leftrightarrow 4)$ & $Y_{6\mu}=\overline{4312}$ \\
\hline
\end{tabular}
\bigskip
\caption{List of elements of $S_4$ ordered as 1+3+8+6+6=24 vector 
bosons, half of which are the Standard Model vector bosons, while the 
rest can be identified with GUT-like X- and Y-bosons. 
In ref. \cite{lampe1} they where 
shown to lead to the correct standard gauge interaction terms. 
The decomposition follows the class structure of the group $S_4$, 
which consists of 5 classes usually called I, $C_2$, $C_3$, 
$C_4$ and $C_2'$ with 1, 3, 8, 6 and 6 elements, respectively. 
In principle one has 2 separate $S_4$ tables, i.e. 2 
separate $S_4$ multiplets, one for 'left' and one for 
'right' vector bosons $V_L$ and $V_R$ which can formally be united in 
one large table by using the octahedral group $O_h \cong S_4 \times P_i$, 
where $P_i$ is an inner parity operation 
defined to transform $V_L \leftrightarrow V_R$. 
Note that the question of (spatial) parity violation 
and vector like interactions has been discussed in ref. \cite{lampe2}.}
\end{center}
\end{table}

Ordering the particle spectra according to representations of the 
permutation group $S_4$ one is naturally lead to the idea 
of a constituent picture where 
fermions (table 1) and gauge bosons (table 2) are built 
from 4 tetrons $t^a$ with 'flavors' $ a \in {1,2,3,4}$ 
and with the condition that in a bound state 
all 4 flavors must be different. The origin of this selection rule 
has been widely discussed in refs. \cite{lampe2,lampe1}. 

The most appealing solution is to allow only discrete 
values for the inner symmetry variable, i.e. 
$t^1$, $t^2$, $t^3$ and $t^4$ are assumed to be fixed 
vectors in the inner symmetry space which point 
to the corners of an inner tetrahedron, 
and then to assume that the interaction 
Hamiltonian is proportional to the volume 
of this tetrahedron. In that case non-permutation 
states like $t^1 t^1 t^1 t^2$ etc are automatically 
suppressed and one ends up with 24 possible bound 
states transforming under representations 
of $S_4$. It may be noted that models with a 
discrete inner symmetry 
space have been extensively studied in the framework 
of lattice physics \cite{seiler,hasenfratz}. 
I will come to a lattice interpretation of 
this point in the appendix. 

Another important question is how the spin-$\frac{1}{2}$ behavior of 
quarks and leptons arise from the spin of the 4 constituents. 
This is the so called 'tetron spin problem' and will 
be discussed next. 


\section{A possible Solution to the Tetron Spin Problem} 

One could have an easy living if one would assume quarks and 
leptons to be composed e.g. of four scalar tetrons and a neutral 
nucleus with spin-$\frac{1}{2}$. 
In the present paper, a different approach will be followed. 

For simplicity, only spatial transformations will be 
considered. Extension to Minkowski space, 
i.e. going from rotational SO(3) to SO(3,1), 
essentially amounts to introduce antitetrons.

Let me start with a few well-known facts about half-integer 
spin: in a physical experiment one cannot distinguish 
between states which differ by a complex phase. Therefore, in 
addition to ordinary representations one may include 
projective, half-integer spin representations of 
the rotation group SO(3), and also of its $T_d\cong S_4$ 
subgroup\footnote{$T_d$ is the rotation symmetry group 
of a regular tetrahedron. It is a subgroup of $O(3)$ 
and isomorphic to $S_4$. 
$T_d$ is also isomorphic to the octahedral 
group $O$, i.e. the group of proper rotations of a 
cube which is a subgroup of SO(3).}. These 
are true representations of the corresponding covering 
groups SU(2) and $\tilde{S_4}$, respectively.

To solve the tetron spin problem I suggest to give up 
the requirement of continuous rotation symmetry and assume that 
tetrons live and interact in microscopical environments, in which 
only permutation symmetry survives. The latter is 
much less restrictive than rotational SO(3), because 
the idea of rotation assumes concepts of angle and length, 
which may be obstacled by quantum fluctuations when approaching the 
Planck scale. In contrast, the idea of permutation merely 
presupposes the more fundamental principle of identity. This is why 
permutation groups may enter theoretical physics at finer 
levels of resolution and higher energies than the Lorentz group. 
Tetrons may be more basic than spinors. 

I call this assumption the 'spatial permutation hypothesis'. 
It amounts to introducing a second permutation index 
called i, j, k or l and taking values 1, 2, 3 and 4 (in 
addition to the tetron 'flavor' index a, b, c and d) 
and being responsible for the 
spatial ('spin') transformation behavior of tetrons 
and its compound states. 

It is true that the phenomenological observation of 24 quarks 
and leptons and their interactions imply a permutation 
principle only on the level of inner symmetries. 
However, the assumption of 4 different tetron 
'spins' within a fermion bound state comes closest to the original 
intuition of a spatial tetrahedral structure as 
discussed in ref. \cite{lampe2} where a generic ansatz 
for the composite wave function $t^{a}_{i} t^{b}_{j} t^{c}_{k} t^{d}_{l}$ 
with $a,b,c,d,i,j,k,l \in \{1,2,3,4\}$ has been proposed. 

As a consequence of the spatial permutation hypothesis 
a new type of particle statistics 
will arise (called {\it tetron statistics}) 
which differs from Fermi and Bose statistics and 
will play a role in the interpretation of the Big Bang and 
cosmic inflation presented below. 

\section{The Details} 

According to the spatial permutation hypothesis, 
the spin part of a 4-particle fermionic compound state 
should transform according to a (projective) representation of $S_4$. 
Besides the ordinary representations $A_1$, $A_2$, $E$, $T_1$ and $T_2$
there are 3 irreducible projective representations (representations 
of the covering group $\tilde{S_4}$), namely 
$G_1$, $G_2$ and $H$ of dimensions 2, 2 and 4, respectively \cite{johnson}. 
The sum 4+4+16 of the dimensions squared accounts for the 24 additional 
elements due to the $Z_2$ covering of $S_4$.
Among them, $G_1$ uniquely corresponds to spin-$\frac{1}{2}$, i.e. is obtained 
as the restriction of the fundamental SU(2) representation to $\tilde{S_4}$. 
Similarly, $H$ can be obtained from the spin-$\frac{3}{2}$ representation 
of SU(2), whereas $G_2$ is obtained from $G_1$ by reversing the sign 
for odd permutations. The combination $G_2+H$ corresponds to a restriction of 
the spin-$\frac{5}{2}$ representation of SU(2) to $\tilde{S_4}$. 

{\footnotesize 

For the understanding of the following arguments a 
short digression on quaternions and its usefulness for 
describing nonrelativistic 
spin-$\frac{1}{2}$ fermions will be helpful: 

Quaternions \cite{conway,kantor,schaf} 
are a non-commutative extension of the complex numbers 
and play a special role in mathematics, because 
they form one of only three finite-dimensional 
division algebra containing the real numbers as a subalgebra. 
(The other two are the complex numbers and the octonions.)
As a vector space they are generated by 4 basis elementes 1, I, J and K 
which fulfill $I^2=J^2=K^2=IJK=-1$, where K can be obtained as a 
product $K=IJ$ from I and J. 
Quaternions are non-commutative in the sense IJ=-JI. 
Any quaternion q has an expansion of the form 
\begin{eqnarray}  
q&=&c_1+J c_2 \nonumber \\
 &=&r_1+Ir_2+Jr_3+Kr_4 
\label{eq302ui}
\end{eqnarray}
with real $r_i$ and complex $c_1=r_1+Ir_2$ and $c_2=r_3-Ir_4$.

There is a one-to-one corresponence between unit quaternions 
$q_0$ and SU(2) transformation  
matrices, because the latter are necessarily of the form 
$(\alpha, \beta; -\beta^*, \alpha^*)$ 
with complex $\alpha$ and $\beta$ fulfilling 
$|\alpha|^2 + |\beta|^2 =1$, and can be written as 
$q_0=\alpha + J \beta$. 
Therefore, the action of SU(2) matrices on spinor fields $(c_1,c_2)$ 
($c_1$ with spin up and $c_2$ with spin down) can 
in quaternion notation be rewritten as: 
\begin{equation} 
c_1+J c_2 \rightarrow (\alpha + J \beta)(c_1+J c_2)
\label{eq20511}
\end{equation}
For example the unit quaternions I and J corresponding to 
rotations by $\pi$ about the x and y-axis amount to  
$c_1 \rightarrow Ic_1, c_2\rightarrow -Ic_2$ 
and $c_1 \rightarrow -c_2, c_2\rightarrow c_1$, 
respectively. 
For a general SU(2) transformation one has  
$c_1 \rightarrow \alpha c_1-\beta^* c_2$ and 
$c_2 \rightarrow \alpha^* c_2+\beta c_1$, 
from which e.g. the antisymmetric 
tensor product combination $c_1 c_2' -c_2 c_1'$ 
can be shown to be rotationally invariant (spin 0). 
} 

To describe spin-$\frac{1}{2}$ bound states one should use 
the symmetry function of the representation $G_1$. This function will 
also be called $G_1$ in the following and can be given 
as linear combination of the $G_1$ representation 
matrices (=unit quaternions): 
\begin{eqnarray}  
G_1&=&  g(1,2,3,4)+  U g(2,3,1,4)+ U^2 g(3,1,2,4)  \nonumber \\
 &+&I g(2,1,4,3)+    S g(3,2,4,1)+ R^2 g(1,3,4,2) \nonumber \\
&+& J g(3,4,1,2)+    R g(1,4,2,3)+ T^2 g(2,4,3,1) \nonumber \\
&+& K g(4,3,2,1)+    T g(4,1,3,2)+ S^2 g(4,2,1,3) \nonumber \\
&+& \frac{I+K}{\sqrt{2}} g(3,2,1,4)+\frac{I-J}{\sqrt{2}} g(1,3,2,4)        
                                     +\frac{J+K}{\sqrt{2}} g(2,1,3,4)  \nonumber  \\
&+&  \frac{1-J}{\sqrt{2}} g(2,3,4,1)+\frac{1-K}{\sqrt{2}} g(3,1,4,2)
                                     +\frac{J-K}{\sqrt{2}} g(1,2,4,3) \nonumber   \\
&+&  \frac{I-K}{\sqrt{2}} g(1,4,3,2)+\frac{1+K}{\sqrt{2}} g(2,4,1,3)
                                     +\frac{1+I}{\sqrt{2}} g(3,4,2,1) \nonumber   \\
&+&  \frac{1+J}{\sqrt{2}} g(4,1,2,3)+\frac{I+J}{\sqrt{2}} g(4,2,3,1)
                                     +\frac{1-I}{\sqrt{2}} g(4,3,1,2)  
\label{eq3029}
\end{eqnarray} 
where $R=\frac{1}{2}( 1- I- J- K), 
S=\frac{1}{2}( 1- I+ J+ K), T=\frac{1}{2}( 1+ I- J+ K)$ 
and $U=\frac{1}{2}( 1+ I+ J- K)$. 
One can see explicitly from this equation, 
which $S_4$ permutation $\overline{ijkl}$ is represented 
in $G_1$ by which quaternion, because the corresponding 
quaternion appears as a 
coefficient of $g(i,j,k,l)$. For example, the permutation $\overline{2341}$ 
is represented by $\pm (1-J)/\sqrt{2}$, and so on. 
In other words, the quaternion coefficients $1, I, J, K, 
(I+K)/\sqrt{2}, ..., R, S, T, ...$ in this equations 
represent the elements of $\tilde{S_4}$\footnote{ 
While $\tilde{S_4}$ itself can be shown to make up the inner shell 
of $D_4$-lattices \cite{dixon}, 
the first half of coefficients in eq. (\ref{eq3029}) represent 
even permutations corresponding to $\tilde{A_4}$ 
which is sometimes called the 'binary tetrahedral group', 
and generates the $F_4$ lattice also called the ring of 
Hurwitz integers (=quaternions with half integer coefficients). 
The Hurwitz quaternions form a maximal order (in the sense of ring theory) 
in the division algebra of quaternions with rational components. This 
accounts for its importance. For example restricting to integer 
lattice points, which seems a more obvious candidate for 
the idea of an integral quaternion, one does not get a maximal order 
and is therefore less suited for developing a theory of left 
ideals as in algebraic number theory. 
What Hurwitz realized, was that his definition of integral quaternions 
is the better one to operate with. 
}. 

Due to the 2-fold covering of $S_4$ each of the 
real functions $g(i,j,k,l)$ 
in eq. (\ref{eq3029}) with its 24 terms is in fact a difference 
\begin{equation} 
g(i,j,k,l)=p(i,j,k,l)-m(i,j,k,l)
\label{eq2g511}
\end{equation}
so as to obtain the 48 terms needed for 
a symmetry function of $\tilde{S_4}$. 

Eq. (\ref{eq3029}) should be considered as the spin factor of the 
4-tetron bound state, whereas the $A_1$, $A_2$, $E$, $T_1$ 
and $T_2$-functions of the ordinary $S_4$ representations 
account for the flavor 
factor. In fact, working out the quaternion multiplications in 
eq. (\ref{eq3029}) and using $K=IJ$ one obtains a representation 
of the form $G_1= c_1+J c_2$ 
with $c_1$ and $c_2$ decribing the 2 spin directions 
of the compound fermions, cf. eq. (\ref{eq20511}). 
Mathematically, the appearance of 2 complex functions $c_1$ and $c_2$ 
in eq. (\ref{eq3029}) is merely expression of the fact that for the 2-dimensional 
representation $G_1$ 4 real(=2 complex) 
symmetry functions can be constructed, which in eq. (\ref{eq3029}) are 
combined in one quaternion function. 

Eq. (\ref{eq3029}) therefore describes a decent fermion state 
which transforms in the standard way, cf. eq. (\ref{eq20511}).
On the other hand, eq. (\ref{eq3029}) also 
inherits the spatial permutation hypothesis (i.e. giving up full 
SU(2) rotational invariance on the tetron level) in 
that the function $G_1$ naturally reacts like a (projective) $S_4$ 
representation under permutations of $i,j,k,l \in \{1,2,3,4\}$. 

The picture followed here is a sort of molecular approach 
where one starts with a fixed 
spatial tetrahedral configuration with 4 distinct 
permutation ('spin') indices $i,j,k,l \in \{1,2,3,4\}$. 
Its reaction under permutations (=tetrahedral $T_d$ transformations) 
of $i,j,k,l$ is dictated by the spatial permutation hypothesis, 
whereas the behavior ($G_1$) under full rotational SU(2) is obtained 
from the requirement that the compound state must be a fermion. 


Since we have given up rotational symmetry on the tetron level, the 
question of how a single tetron $t_i^a$ with index $i \in \{1,2,3,4\}$ 
transforms into itself under rotations 
need not be discussed. It is merely 
necessary to know how compound states transform unter permutations of 
indices $i,j,k,l \in \{1,2,3,4\}$ and this question is 
answered by the symmetry function $G_1$.

In other words: since $G_1$ is not contained in any 
tensor product of 4 $S_4$ representations, one can only interpret 
the inner symmetry part of the wavefunction as a tensor product 
of tetron factors, but not the 
functions $g(i,j,k,l)$ or p and m of eq.(\ref{eq2g511}). 
\begin{equation} 
g(i,j,k,l) \neq t_{i}\otimes t'_{j}\otimes t''_{k}\otimes t'''_{l}
\label{eq302rr}
\end{equation} 
Nevertheless, I will sometimes use the tensor notation 
for the sake of illustration. For instance, 
the complete 'spin' and 'flavor' wave function of 
quarks and leptons can then plainly be denoted as
\begin{eqnarray}  
t^{a}_{1}\otimes t^{b'}_{2}\otimes t^{c''}_{3}\otimes t^{d'''}_{4}  &+& t^{b}_{1}\otimes t^{c'}_{2}\otimes t^{a''}_{3}\otimes t^{d'''}_{4} + ... \nonumber \\
I t^{a}_{2}\otimes t^{b'}_{1}\otimes t^{c''}_{4}\otimes t^{d'''}_{3} &+& ... \nonumber \\
...& &
\label{eq302}
\end{eqnarray}
Here in the rows the tetron flavor indices $a,b,c,d$ are permutated in order 
to obtain the appropriate flavor combination ($A_1$ of $S_4$ as an example, 
for the $A_2$, $T_1$ etc flavor representations $G_2$ and $H$ will 
come into play), 
whereas in the columns the tetron 'spin' indices $i,j,k,l$ are permutated 
in order to obtain the $G_1$ spin combination.\footnote{
Note that in general, the permutation of the tensor product 
indices - denoted by primes in eq. (\ref{eq302rr}) - must not be 
messed up with the permutation of spin states. Only 
in the case at hand, where 4 different spin states in 
4 different tensor factors are considered, there is no difference. 
}

In summary, 
eq. (\ref{eq3029}) should be considered as the spin factor of the 
quark and lepton states, whereas the $A_1$, $A_2$, $E$, $T_1$ 
and $T_2$-functions of the ordinary $S_4$ representations 
account for the inner symmetry 'flavor' factor. (Those 
functions can be found, for example, in ref. \cite{lampe1}.)
The full quark and lepton spectrum of table 1 including spatial and 
inner symmetries can then be written as 
\begin{equation} 
(A_1 + A_2 +2E+3T_1+3T_2)_{in}\otimes G_{1sp}
=24 G_1 
\label{eq17udz}
\end{equation}
where $in$ stands for the inner and $sp$ for the spatial 
part of the wave function, 
and the factor of 24 on the r.h.s. accounts for the 
24 degrees fo freedom of 3 fermion families.\footnote{There 
is a pictorial interpretation of the fermion bound state 'molecules', 
where the tetrahedron as a whole forms a sort of molecular 
cluster, and the 24 inner $S_4$ symmetry configurations 
can be thought to be realized in ordinary space. 
Namely, on each of the 4 corners of a tetrahedron a 
single tetron $t^\alpha_\iota$ is located which is 
composed of a 'nucleus' $\alpha \in {1,2,3,4}$ surrounded 
by a 'cloud' $\iota \in {1,2,3,4}$. 
Inner symmetry transformations act by interchanging 
the clouds whereas under spatial rotations 
both nuclei and clouds are transformed simultaneously. 
In other words, the 24 flavor states 
$(A_1 + A_2 +2E+3T_1+3T_2)_{in}$
can be obtained by varying $\iota$ for fixed $\alpha$,  
whereas unter spatial tetrahedral transformations 
the $G_{1sp}$-combination of indices should be chosen 
with varied $\iota$ and $\alpha$ simultaneously. 
} 


\section{Dark Matter from Tetrons}

Dark matter is a hypothetical type of matter that is
undetectable by its emitted radiation, but which can be 
inferred only from gravitational effects. Its presence is
postulated to explain the flat rotation curves of spiral galaxies and
other evidence of missing mass in the universe. 
According to present observations, there exists between 4 and 6 times 
more dark matter than ordinary matter in the universe. Further 
it is known, that it must be composed of mostly 
cold, i.e. nonrelativistic, particles.  

We have seen in the last section, how the spin-$\frac{1}{2}$ nature 
of quarks and leptons can be deconstructed using the $G_1$ representation 
of the permutation group. It is certainly true that 
the phenomenological observation of 24 quarks 
and leptons and their interactions suggests a permutation 
principle only on the level of {\it inner} symmetries. However 
due to the problems which arise in connection with spin and 
statistics we were naturally lead to consider the possibility 
that there is a spatial $S_4$-index as well 
and that this can be used to understand the spin-$\frac{1}{2}$ 
nature of quarks and leptons. 

In the following I want make use 
of this procedure to show that there are natural dark matter 
candidates in the tetron model responsible 
for the bulk of the observed dark matter in the universe. 
Namely, if this approach has some meaning it is tempting that 
besides $G_1$ also the two other half-integer spin 
representations of $\tilde{S_4}$ ($H$ and $G_2$) play 
a role in nature, or in other words, that together 
with ordinary ($G_1$-)matter sets of particle families with 
spin $\frac{3}{2}$ ($H$) and spin $\frac{5}{2}$ ($G_2+H$) should 
have been produced during cosmogenesis. In fact, 
eq. (\ref{eq17udz}) naturally extends to 
\begin{equation} 
(A_1 + A_2 +2E+3T_1+3T_2)_{in}\otimes (G_1+G_2+2H)_{sp}
=24(G_1+G_2+2H) 
\label{eq11dz}
\end{equation}
As before, $in$ stands for the inner and $sp$ for the spatial 
$S_4$ index set 
and the factor of 24 on the r.h.s. accounts for the 
24 'flavor' degrees of freedom of 3 times 3 fermion families 
for $G_1$, $H$ and $G_2+H$ each with particle masses of 
roughly comparable size. 

Next, it will be assumed that - apart from gravity forces - the 
new ($G_2$ and $H$) fermions decouple from 
ordinary ($G_1$) fermions, 
i.e. that spin-$\frac{3}{2}$ and spin-$\frac{5}{2}$ matter have 
interactions completely separate from those of ordinary 
matter.\footnote 
{
It is an interesting question how the interactions 
among the dark matter ($G_2$ and $H$) fermions look like 
and whether they lead to atomic and molecular binding 
states similar to what we are used from ordinary matter or whether 
the spin-$\frac{3}{2}$ and spin-$\frac{5}{2}$ quarks will 
not be confined and exist as free particles. A natural ansatz 
is to extend the vector boson content of table 2 in a manner 
compatible with permutation symmetry. In fact one may summarize 
the content of table 2 as 
\begin{equation} 
(A_1^{\pm} + A_2^{\pm} +2E^{\pm}+3T_1^{\pm}+3T_2^{\pm})_{in}\otimes T_{1sp} 
\label{eq15tt}
\end{equation} 
in terms of $O_h$ representations $R^{\pm}$, where $+$ stands for 
lefthanded and $-$ for righthanded vector bosons. 
One may try to extend this expression to include the 
interactions among the $H$- and $G_2+H$-fermion families: 
\begin{equation} 
(A_1^{\pm} + A_2^{\pm} +2E^{\pm}+3T_1^{\pm}+3T_2^{\pm})_{in}
\otimes (A_1 + A_2 +2E+3T_1+3T_2)_{sp}
\label{eq15ttt}
\end{equation}
and has to show that the additional bosons interact only 
within the $H$- and $G_2+H$-fermion families, 
but not with ordinary (i.e. $G_1$-) matter. 
}

Assuming further, that initially all matter fields 
are produced at uniform rates, one expects a ratio 
of 1:5 for the relative distribution of matter (including 
neutrinos) and dark matter in the universe. 
This ratio is obtained by counting the spin degrees of freedom 
2:(4+6) of spin-$\frac{1}{2}$, -$\frac{3}{2}$ and -$\frac{5}{2}$ objects 
or equivalently from the ratio of dimensions $dim(G_1):dim(G_2+2H)$ 
and should be considered as one of the main results of the 
present paper. 
The fact that only 3 representations are involved has to 
do with the fact that $S_4$ is a finte group with a finite 
number of representations. 

The idea behind this consideration is, that 
at Big Bang energies where masses play no role, all 3 matter 
types ($G_1$, $G_2$ and $H$) are produced in equal amount corresponding 
to a mass energy ratio of ordinary to dark matter 
$dim(G_1):dim(G_2+2H)=1:5$ and that this ratio has not 
changed since that time because apart from gravity there are no interactions 
between the 3 matter types. In other words, all decays and transitions 
take place only within one of the matter types and do not disturb the 
ratio 1:5. The same holds true for radiation: when an 
electron-positron pair annihilates, 
a photon of type $G_1$ appears, and this can only annihilate into a 
fermion-antifermion pair of type $G_1$. The reason for this 
lies in the manner in which the photon - and also the gluon and 
the W-boson - are constructed in the tetron model as 
scattering states of $G_1$-fermions only \cite{lampe1}.


\section{A new Statistics}

Eq. (\ref{eq3029}) reflects the statistical behavior of a  
4-tetron conglomerate under permutations of its components. 
This behavior has a 
certain similarity to that of fermions but is certainly not 
identical. While conglomerates of fermions usually transform 
with the totally antisymmetric representation (like $A_2$), 
tetrons go with $G_1$, which gives a factor of I under 
the exchange ($1\leftrightarrow 2,3\leftrightarrow 4$)
or $\frac{1}{\sqrt{2}} (J+K)$ under ($1\leftrightarrow 2$), 
whereas a 2-fermion conglomerate in a $A_2=c_1 c_2' -c_2 c_1'$ 
configuration responds with -1 (i.e. antisymmetric) to the exchange 
of ($1\leftrightarrow 2$). See table 3, where the behavior of 
tetrons and fermions is compared. The fact that tetrons behave 
more complicated under transpositions $(i \leftrightarrow j)$, 
has to do with the fact that transpositions in $S_4$ correspond 
to relatively complicated space transformations in $T_d$. 

We therefore conclude that tetrons follow their own statistics 
which is neither bosonic nor fermionic, and assert, that 
a sort of 'tetron spin statistics theorem' holds, which 
allows only bound states in which all tetron flavors are different 
(cf. the selection rule / exclusion principle mentioned at the end 
of section 2).

\begin{table}
\label{tab4}
\begin{center}
\begin{tabular}{|c|c|}
\hline
FERMIONS                             &  TETRONS   \\
\hline
\hline
\multicolumn{2}{|c|}{compound states:}                         \\
\hline
boson from 2 fermions:                &  fermion from 4 tetrons: \\
complex tensor product                &  quasi-complex, quaternion tensor product \\
$A_2=c_1 c_2' -c_2 c_1'$              &  $G_1=g(1,2,3,4)+I g(2,1,4,3)+J...$ \\   
                                      &  $=t^{a}_{1} t^{b'}_{2} t^{c''}_{3} t^{d'''}_{4} + I t^{a}_{2} t^{b'}_{1} t^{c''}_{4} t^{d'''}_{3} +...$ \\
bosonic behavior under rotations      &  fermionic behavior under rotations \\
                                      &  $G_1\rightarrow (\alpha + J \beta)G_1$ \\
\hline
\hline
\multicolumn{2}{|c|}{permutation behavior/statistics:}                         \\
\hline
-1 under ($1\leftrightarrow 2$)      &   a factor I under ($1\leftrightarrow 2,3\leftrightarrow 4$) \\  
                                     &   a factor $\frac{1}{\sqrt{2}} (J+K)$ under ($1\leftrightarrow 2$) etc \\
\hline
\end{tabular}
\bigskip
\caption{Comparison between the known fermion behavior and the anticipated 
tetron behavior.}
\end{center}
\end{table}


\section{Gravitons, Quintessence and the Interaction among Tetrons}

In this section I follow the idea that the gravitational field 
can be described in terms of tetron constituents. 
This could be either in the form of a van-der-Waals remnant of 
the interactions among tetrons or, in more concrete terms, 
of a composite gravitational field, described in terms 
of tetron interactions. 

What is the possible form of the interaction among tetrons? 
On an effective Lagrangian level it involves 
4-tetron product terms like $t^a_{i} t^b_{j} t^c_{k} t^d_{l}$. 
It would be desirable to interpret this as 
an effective interaction which can be traced back to an interaction 
of 2 tetrons of the form $t^a_{i} t^b_{j} B_{kl}^{cd}$, 
with $i,j,k,l,a,b,c,d \in \{1,2,3,4\}$ and 
$B_{kl}^{cd}$ being some interaction 'field'. 
Note that as before, 
no specific spatial transformation properties 
can be assigned to a single index i or j. 
However, in the combination $ijkl$ they 
will transform under an $S_4$ representation. 

Since gravity is flavor independent, in order 
to construct it from B-fields, these must not 
depend on the flavor indices a, b. Therefore the 2-tetron 
interaction simplifies to
\begin{equation} 
L_{ttB}=t^a_{i} t^b_{j} B_{kl} 
\label{eq123uu}
\end{equation}
In pictorial language the B-field occupies 
the 6 edges of a tetrahedron. 

In concrete terms the graviton will be assumed to be a bound state of 
two B-fields. Furthermore it should meet the general 
selection rule / exclusion principle formulated in ref. \cite{lampe1} 
that every physical field must be 
a permutation field. Then - in the same way as fermion states 
were written down with the help of the representation 
$G_1$ eq. (\ref{eq3029}) - the gravitational field can 
be expanded with the help of spin-2 representation matrices 
$R_{\mu \nu}(ijkl)$ of $S_4$ 
given by the representation $E+T_2$ of $S_4$: 
\begin{eqnarray}  
g_{\mu \nu} = R_{\mu \nu}(1234) B_{12} B_{34} + 
R_{\mu \nu}(2143) B_{21} B_{43} + ...
\label{eq3dd}
\end{eqnarray} 

In the following explicit construction the spin-2 
representation will be formally calculated from a product 
of 2 vector representations 
\begin{equation} 
T_1 \otimes T_1=A_1+T_1+E+T_2
\label{eq33dz}
\end{equation}
of the spatial $S_4$-symmetry indices, 
where $A_1$, $T_1$ and $E+T_2$ represent the spin-0, spin-1 
and spin-2 contributions to the product, respectively. 
Furthermore, the temporal gauge $g_{0\mu}=0$ will be used 
which, at least in the weak field approaximation, 
is known to be compatibel with the harmonic gauge 
often used in relativistic calculations \cite{mag}. 

The metric tensor then takes the form 
\begin{equation} 
g_{\mu\nu}=
\left(\begin{array}{cccc}
-t_{XX}-t_{YY}-t_{ZZ} & 0 & 0 & 0  \\
0 & t_{XX} & t_{XY}  & t_{XZ} \\
0 & t_{YX} & t_{YY}  & t_{YZ} \\
0 & t_{ZX} & t_{ZY}  & t_{ZZ} 
\end{array}\right)
\label{eq35dz}
\end{equation} 
Here we have allowed for a nonvanishing $g_{00}$ contribution 
due to the singlet $A_1$ which may represent the 
quintessence scalar $\phi_q$ \cite{wetterich} 
appearing in solutions to the dark energy 
problem and a possible antisymmetric component of $g_{\mu\nu}$ 
stemming from the spin-1 contribution $T_1$ on the r.h.s. of
eq. (\ref{eq33dz}). The antisymmetric components may play 
a role in the so-called scalar-vector-tensor model 
\cite{bekenstein} and in gravity with torsion \cite{cartan}. 
Making use of the appropriate Clebsch-Gordon 
coefficients \cite{griff} 
the relation of $g_{\mu\nu}$ eq. (\ref{eq35dz}) to the known 
$S_4$ representation matrices \cite{lampe1} is given by 
\begin{eqnarray}  
A_1&=&t_{XX}+t_{YY}+t_{ZZ} \\ 
E_{11}&=&(t_{XX}-t_{YY})/2 \\
E_{12}&=&(t_{XX}+t_{YY}-2t_{ZZ})/\sqrt{6}  \\
T_{2,11}&=&(t_{XY}+t_{YX})/2 \\
T_{1,11}&=&(t_{XY}-t_{YX})/2 
\label{eq3vv}
\end{eqnarray} 
etc. Putting everything together, one obtains for example for the 
($A_1$) quintessence field 
$\phi_q=\sum_{ijkl} B_{ij} B_{kl}$ where the sum runs over all 
permutations $\overline{ijkl} \in S_4$. Similarly for the 
spatial components of $g_{\mu\nu}$ from the other 
representations $E$, $T_1$ and $T_2$. 

It should be noted that, instead of using 2-B-field bound states, 
in the lattice interpretation given in the appendix one may be 
more general and assume that 
the graviton and its companions are excitations within 
the permutation lattice of the general form 
\begin{equation} 
1_{in} \otimes (A_1 + A_2 +2E+3T_1+3T_2)_{sp}
\label{eq48dz}
\end{equation}
Since - in contrast to eq. (\ref{eq11dz}) - there is no inner 
symmetry index, only one $A_1$, one $A_2$, one $T_1$ and 
one $E+T_2$ field emerge on the ground state level. 
This corresponds to a scalar field $\phi_q$, an axial scalar 
$\phi_a$, a spin-1 vector $U_{\mu}$ and a spin-2 tensor field. 
In the massless limit 
the transversal modes of the spin-1 and spin-2 
excitations will vanish and a graviton and a vector field each with 
2 helicities appear. 

Having constructed the compound states one can try to write 
down their effective interactions. The requirement of local 
Lorentz invariance more or less fixes the Lagrangian to be 
\cite{wetterich,bekenstein,farrar,mag}
\begin{eqnarray}  
L = \frac{1}{2} \sqrt{-g}  M_P^2 R + 
L(\phi_q) + L(\phi_a) + L(U_{\mu}) +L_{WW}
\label{equzt}
\end{eqnarray} 
where R is the Ricci scalar associated with the graviton, g is 
the determinant of the (symmetric) metric tensor 
and $M_P = 1/\sqrt{8\pi G}$ the reduced Planck mass. 
\begin{eqnarray}  
L(\phi_q) = \frac{1}{2} \partial_{\mu}\phi_q \partial^{\mu}\phi_q 
            -V(\phi_q) 
\label{equztqq}
\end{eqnarray} 
denotes the quintessence part of the Lagrangian 
\cite{wetterich,farrar,mag}. Similarly 
for $L(\phi_a)$ and $L(U_\mu)$, whereas 
$L_{WW}$ denotes interactions among the various 
fields\cite{bekenstein}.

Exploring the phenomenology of eqs. (\ref{equzt}) and 
(\ref{equztqq}) requires a form for the potential $V(\phi_q)$. 
In order to account for the dark energy component of the 
total cosmic mass energy, 
this is usually chosen in such a way that the field stress-energy
tensor approximates the effect of a cosmological
constant\cite{wetterich,farrar,anderson}.


\section{A Tetron Plasma in the very early Universe}

According to the cosmological Standard Model the universe 
began in a state, in which spacetime and physical laws have no 
real meaning. This so called Planck era lasted about 5,4 10$^{-44}$ s. 
Only after that spacetime and 
matter came into being in the process of the Big Bang, 
and the laws of physics came into action. 

As discussed in section 4 the question how a single tetron 
behaves under spacetime transformations is not well put, because 
single tetrons cannot be isolated spatially. 
Therefore I want to develop a picture that in the Planck era 
the universe consisted of a countable set of a large number 
of tetrons and B-fields - just a set, with no spatial properties, 
but possibly with interactions governed by permutation symmetry - 
and that the physical history of the universe as a spacetime manifold 
began only, when the tetrons formed $S_4$ bound states, which transform under 
representations of the rotation or Lorentz group. 

I will call the state before the advent of bound states 
a tetron plasma - although one may object that 'plasma' is perhaps 
not the right word for a set of tetrons without 
a metric space, so that for example particle velocities, 
energies and probably even temperature cannot be defined.
As a countable, practically infinite set it has a 
$S_\infty$ permutation symmetry, which in the process of 
bound state formation gets broken to $S_4$. About 
the nature of the symmetry breaking $S_\infty \rightarrow S_4$ 
one can only speculate. It may have to do with 
Bott periodicity which honours spatial dimensions 
of 3 and 7, because in these dimensions division algebra 
structures can be imposed on the corresponding 
vector spaces (cf. the discussion at the end of 
the appendix). 

The appearance of $S_4$-symmetry and of a 3-dimensional 
space are actually correlated. 
Namely, according to 
eqs. (\ref{eq3dd}) and (\ref{eq3029}) 4-tetron and 
2-B-field aggregates constitute points in space by defining a 
transition from a set on which only permutation operations 
act ($ijkl$... $\rightarrow$ permutations of $ijkl$...) to a continuus 
space where rotational symmetry transformations 
$R_{\mu\nu}(ijkl)$ (eq. (\ref{eq3dd})) and $I,J,K$ etc 
(eq. (\ref{eq3029})) are defined. 
In that sense it may be said that tetron interactions 
constitute 3-dimensional space. 

The transition of the universe from a tetron plasma to 
the later radiation and matter phases could be related 
to cosmic inflation, because a 
tetron plasma governed by tetron statistics during the 
Planck era could account for the pressure 
required in the inflationary scenario, by means of 
the enormous bindung energies set free 
when quarks, leptons and radiation states are formed and space 
is blown up from a discrete set(=tetron plasma), where 
distances are not defined to a curved semidiscrete manifold, 
where the extension of bound states is roughly given 
by the Planck scale. 

Unfortunately, in its present stage the tetron model 
does not provide a suitable dynamical scheme 
and therefore does not have enough quantitative 
predictive power to compete with current Lagrangian 
approaches \cite{bassett,barvinsky,bezrukov} to inflation. 
In the Lagrangian models the effects of inflation are 
described by an (effective) Lagrangian containing a 
scalar inflaton field with a definite dynamics, 
which is able to quantitatively explain the mechanism 
which drives the rapid expansion in the inflation period 
(for a comprehensive review see e.g. ref. \cite{bassett}). 
This field may or may not be one of the Higgs 
fields appearing in standard particle physics models 
and with minimal \cite{barvinsky} or 
non-minimal \cite{bezrukov} coupling to gravity. 
Prior to the expansion period, the inflaton is at a 
higher energy state. A suitable potential or 
random quantum fluctuations then generate a repulsive force and 
trigger a phase transition whereby the inflaton 
field releases its potential energy as matter 
and radiation as it settles to its lowest energy state. 

The Lagrangian approach to inflation can be interpreted 
in different ways. One way (preferred by the present 
author) is to argue that these models (of which there 
are hundreds) provide a convenient method of parametrizing
the early universe but that, because they
are fundamentally semi-classical, are unlikely to be a true
description of the physics underlying the very early universe.
The other (probably more common) approach is
to argue that the inflaton is the true source of
inflation and that its identity may be found by considering 
one of the extensions of the standard model
based on grand unified theories, supergravity
or string theory, from which then  
definite quantitative predictions can be obtained. 

In comparison, the tetron model arguments in favor of 
inflation are only qualitative in nature. 
Nevertheless, they may lead to a microscopic understanding 
of an effective inflaton interaction, 
once a model for the dynamical behavior of tetrons 
and of a tetron plasma is developed. 


\section{Summary}

In summary, the tetron model modifies 
the standard Big Bang scenario in various respects. Prior 
to the epoche of radiation, quark-gluon plasma etc governed by 
GUT or Standard Model interactions 
there may a tetron plasma governed by tetron statistics, 
where distances, angles and a metric 
do not exist but arise only when tetronic bound states 
are formed. 
The formation of these bound states sets 
free an an enormous amount of binding energy and 
introduces the pressure needed for inflation 
in the early universe. 

Furthermore, it was shown how the tetron model 
yields the physical particles, i.e. fermions of the 
form $\sim t^4$, radiation $\sim (\bar t t)^4$ and 
gravitational interactions $\sim B^2$ 
as well as more speculative fields of 
spin $\frac{3}{2}$ and spin $\frac{5}{2}$ 
which may serve as dark matter candidates. 

There are several objections which may be raised 
against the tetron model. 
One is that at its current state it relies mainly 
on group theoretical arguments and not much can be said 
about the dynamical behavior of tetrons. 
What seems to be certain, however, is that due to the 
discrete-like features of the model the ultraviolet 
treatment of the tetron theory will be quite 
different from the renormalization one usually 
encounters at small distances in quantum field theories. 

Further, one could suspect that the tetron model contradicts 
the Weinberg-Witten theorem \cite{weinberg} which states that no 
massless (composite or elementary) particles with
spin greater than one are consistent with any renormalizable
Lorentz invariant quantum field theory (excluding only
nonrenormalizable theories of gravity and supergravity). 
However, in the case at hand the theorem does 
not apply, because we have abandoned 
Lorentz symmetry from the start replacing it 
by the spatial permutation hypothesis (cf. section 3) 
or by a spatial 'permutation' lattice (cf. the appendix). 



\section{Appendix: An alternative Approach using Lattices} 

The tetrahedral or permutation group $S_4$ is not only 
the symmetry group of a regular tetrahedron, but also of 
a tetrahedral lattice or of a fluctuating $S_4$-permutation 
lattice. 

It is therefore tempting to assume, that the 
inner symmetry space of tetrons is not 
continuous (with a continuous symmetry group) but 
has instead the discrete structure of such a tetrahedral 
lattice. The observed quarks and leptons can then be 
interpreted as excitations on this lattice and 
characterized by representations of the lattice symmetry 
group $S_4$, i.e. by $A_1 + A_2 +2E+3T_1+3T_2$, 
just as in the 'classical' tetron model presented 
in the main text. 
In this picture the original dynamics is governed 
by some unknown lattice interaction 
instead of by four real tetron constituents.  

The lattice ansatz also naturally explains the selection 
rule mentioned in section 2 (that all physical states must be 
permutation states), just because only representations 
of the permutation group $S_4$ are allowed. 

In the following I will make the additional assumption 
that not only the inner symmetry is discrete 
but that physical space is a lattice, too. 
The reason for this is that 
although theories with a discrete inner symmetry over 
a continuous base manifold have been examined \cite{belavin} 
they seem to me rather artificial because they 
will usually lead to domain walls and other discontinuities. 

More precisely, in the spirit of the spatial 
permutation hypthesis (section 3) 
instead of a fixed spatial lattice the existence of a 
spatial permutation lattice with symmetry group $S_4$ 
will be assumed where 
the lattice points are not a priori fixed but may be 
fluctuating due to quantum effects. 
The lattice spacing would be typically 
of the order of the Planck scale with 
the extension of the bound states slightly larger. 


One could ask, why the (inner) lattice structure is seen in the 
flavour spectrum part of eq. (\ref{eq17udz}) whereas the spatial 
part $G_{1sp}$ to a human observer 
appears as spin-$\frac{1}{2}$ representations 
of the {\it continuous} rotation group. The point is that 
with respect to the spatial lattice present physical 
experiments always work at distances much larger than the 
lattice spacing ($\cong M_P$) 
whereas for the inner symmetry lattice we do 
{\it not} encounter the continuum limit, so that the 
representations $A_{1,2}$, $E$ and $T_{1,2}$ remain relevant 
for the particle spectrum.  

A drawback of the lattice picture 
as compared to the tetron constituent 
model, is that it is still less specific and 
there is a larger amount of arbitrariness 
concerning the origin of the observed spectrum 
$(A_1 + A_2 +2E+3T_1+3T_2)_{in}$ for 
quarks and leptons. 
These are interpreted as lattice excitations, and one may, 
for example, assume the existence of 'elementary' excitations 
$g_{1 in}$, $t_{1 in}$ and $h_{in}$ on the inner symmetry lattice 
(transforming with respect to the representations $G_1$, $T_1$ and $H$, 
respectively) from which the quark and lepton spectrum is built 
according to 
\begin{eqnarray}  
g_{1 in} \otimes t_{1 in} \otimes h_{in} = (A_1 + A_2 +2E+3T_1+3T_2)_{in}
\label{equu2ui}
\end{eqnarray}
However, apart from the fact that there are other 
possibilities of tensor products that yield the same result, 
the physical meaning of the 'elementary' excitations is 
rather unclear. 

There is a similar situation in the spatial sector of the 
model, where one would like to obtain the 
spectrum $(G_1+G_2+2H)_{sp}$ from some elementary 
tensor product. 
Again, it turns out that there are several 
different combinations of elementary 
excitations leading to the required result. 

One may speculate whether 
a unification of the spatial and inner symmetry 
sector could remedy the arbitrariness. 
What I have in mind is a compactification scenario 
where one starts with a n-dimensional lattice 
(or n+1 in a relativistic scenario to include a time variable), 
n-3 dimensions of which being compactified. 
The most natural choice seems to be n=7 
because it allows spinorial structures 
which is inherited to the n=3 base manifold 
in the process of compactification.
Due to lack of time I have not yet analyzed 
this promising possibility in detail. 


\end{document}